\documentclass[preprint,prl,epsf,aps]{revtex4}
\usepackage{bm}
\input epsf

\begin{document}

\title{ Spin-galvanic effect due to optical spin orientation}

\author{S.D.~Ganichev$^{1,2}$, Petra~Schneider$^1$, V.V.~Bel'kov$^2$,
E.L.~Ivchenko$^2$, S.A.~Tarasenko$^2$,  W.~Wegscheider$^1$, D.~Weiss$^1$,
D.~Schuh$^3$, B.N.~Murdin$^4$, P.J.~Phillips$^5$, C.R.~Pidgeon$^5$, D.G.~Clarke$^4$,
M.~Merrick$^4$, P.~Murzyn$^5$, E.V.~Beregulin$^2$, and  W.~Prettl$^1$}
\address{$^1$~ Fakult\"{a}t
Physik, University of
Regensburg, 93040, Regensburg, Germany\\
$^2$~A.F.~Ioffe Physico-Technical Institute,
194021 St.~Petersburg, Russia\\
$^3$~Walter Schottky Institute, TU Munich,  D-85748 Garching, Germany\\
$^4$University of Surrey, Guildford, GU2 7XH, UK\\
$^5$Department of Physics, Heriot-Watt University, Edinburgh, UK}

\date{\today}

\begin{abstract}

Under oblique incidence of circularly polarized infrared radiation
the  spin-galvanic effect has been unambiguously observed in
(001)-grown $n$-type GaAs  quantum well (QW) structures in the
absence of any external magnetic field. Resonant inter-subband
transitions have been obtained making use of the tunability of the
free-electron laser FELIX. It is shown that a helicity dependent
photocurrent along one of the $\langle 110 \rangle$ axes is
predominantly contributed by the spin-galvanic effect while that
along the perpendicular in-plane axis is mainly due to the
circular photogalvanic effect. This strong non-equivalence of the
[110] and [1$\bar{1}$0] directions is determined by the interplay
between bulk and structural inversion asymmetries. A microscopic
theory of the spin-galvanic effect for direct inter-subband
optical transitions has been developed being in good agreement
with experimental findings.
\end{abstract}


\maketitle \draft

\newpage

The spin of electrons and holes in solid state systems is an
intensively studied quantum mechanical property showing a large
variety of interesting physical phenomena. Lately there is much
interest in the use of the spin of carriers in semiconductor
heterostructures together with their charge for novel applications
like spintronics~\cite{spintronicbook02}. The necessary conditions
to realize spintronic devices  are high spin polarizations in low
dimensional structures and large spin-splitting of subbands in
{\boldmath $k$}-space. The latter is important for the ability to
control spins with an external electric field by the Rashba
effect~\cite{Bychkov84p78}. Significant progress has been achieved
recently in generating large spin polarizations, in demonstrating
the Rashba splitting and also in using the splitting for
manipulating the spins~\cite{spintronicbook02}.
At the same time
as these conditions are fulfilled it has been shown that the spin polarization
itself drives a current if
the spins are oriented  in the plane of the quantum well (QW)~\cite{Nature02}.
This spin-galvanic effect
was previously demonstrated with optical excitation and the
assistance of an external magnetic field to achieve an in-plane
polarization. As a step towards the long term aim of showing its
existence with only electric injection we report here the
demonstration of the optically induced spin-galvanic effect in
zero magnetic field. We also present the microscopic theory of this effect.

\begin{figure}[h]
   \centerline{\epsfxsize 60mm \epsfbox{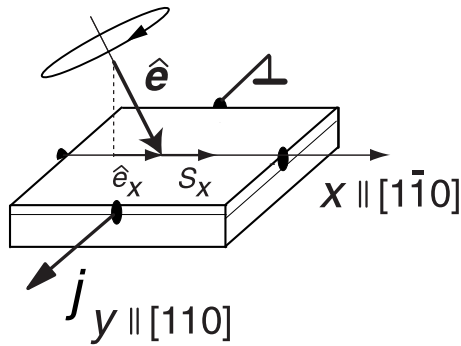}}

\caption{ Geometry of the experiment. At  oblique  incidence of
radiation we obtained projections on the $x$ or $y$ directions of
the unit vector {\boldmath$\hat{e}$} and the averaged spin
 {\boldmath$S$}. The current {\boldmath$j$} is recorded
perpendicular to the direction of light propagation. }
\label{fig1}
\end{figure}

The spin-galvanic effect has been observed at  room temperature by
studying transitions between size quantized subbands $e1$ and $e2$
in $n$-type GaAs/AlGaAs quantum wells (QW). Typical samples, grown
along $z$ $\parallel$ [001] by molecular-beam-epitaxy, consisting
of 30~QWs with a well width of 7.6~nm, 8.2~nm and 8.6~nm, and
free-carrier density in a single well $n_e$ of about
$2\cdot10^{12}$\,cm$^{-2}$ were investigated at room temperature.
Samples were quadratic in shape, with edges oriented along the $x$
$\parallel$ [1$\bar{1}$0] and $y$ $\parallel$ [110]
crystallographic directions. Two pairs of ohmic contacts were
attached in the center of opposite sample edges (see
Fig.~\ref{fig1}).

\begin{figure}[h]
   \centerline{\epsfxsize 86mm \epsfbox{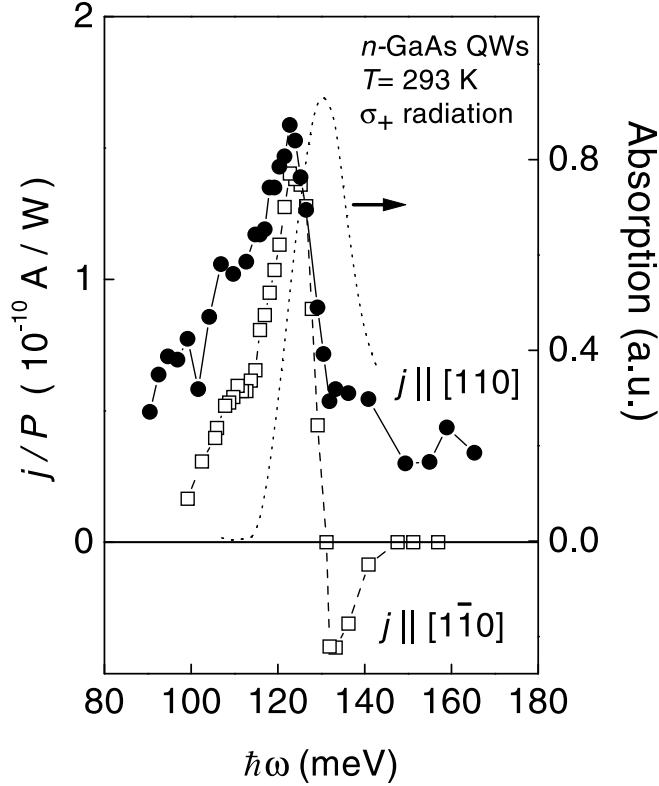}}

\caption{ Photocurrent in QWs normalized by the  light power $P$
at oblique incidence of right-handed circularly polarized
radiation on $n$-type (001)- grown GaAs/AlGaAs QWs of 8.2~nm width
at $T=~$293~K as a function of the photon energy $\hbar \omega$.
Circles: current in [110] direction in response to irradiation
parallel [1$\bar{1}$0]. Squares:  current in [1$\bar{1}$0]
direction in response to irradiation parallel [110]. The dotted
line shows the absorption measured using a Fourier transform
spectrometer.} \label{fig2}
\end{figure}

In Fig.~\ref{fig2} the absorption spectrum of  the sample
containing 8.2~nm wide QWs obtained by Fourier transform
spectroscopy in a waveguide geometry is shown by the dotted curve.
The $e1$ to $e2$ resonance occurs at the photon energy 130.4~meV and
the full width half maximum was 16.8~meV. In order to excite
resonantly and to obtain a measurable photocurrent it was
necessary to have a tunable high power radiation source for which
we used the free electron laser ''FELIX'' at FOM-Rijnhuizen in The
Netherlands~\cite{Knippels99p1578}. The output pulses of light
from FELIX were chosen to be 3 ps long, separated by 40~ns, in a
train (or ''macropulse'') of duration of 5~$\mu$s. The macropulses
had a repetition rate of 5~Hz.

On illumination of the QW structures  by circularly polarized
radiation  at oblique incidence in ($xz$)- or ($yz$)-plane  a
current signal perpendicular to the plane of incidence was
measured, e.g. in $y$ direction for the configuration depicted  in
Fig.~\ref{fig1}. Left handed ($\sigma_-$)  and right handed
($\sigma_+$) circularly polarized radiation was achieved making
use of a Fresnel rhomb. The photocurrent signals generated in the
unbiased devices at room temperature were measured via an
amplifier with a response time of the order of 1 $\mu$s, i.e.
averaged over the macropulse. The voltage in response to a laser
pulse was recorded by an oscilloscope.

The observed current is  proportional  to the helicity $P_{circ}$
of the radiation. The photon energy dependence of the current was
measured for incidence in two different planes with in-plane component of
propagation along the $x$ and $y$ directions. In Fig.~\ref{fig2}
the observed current for both directions is plotted as a function of
photon energy $\hbar \omega$ for $\sigma_+$ polarized radiation
together with the absorption spectrum. It can be seen that for
current along $x$ $\parallel$ [1$\bar{1}$0] the shape is similar
to the derivative of the absorption spectrum, and in particular
there is a change of sign which occurs at the line center of the
absorption. When the sample was rotated by 90$^\circ$ about $z$,
so that light propagates now along $x$ and the current flows along
$y$ $\parallel$ [110], the sign reversal in the current disappears
and its shape follows more closely the absorption spectrum.

\begin{figure}
   \centerline{\epsfxsize 120mm \epsfbox{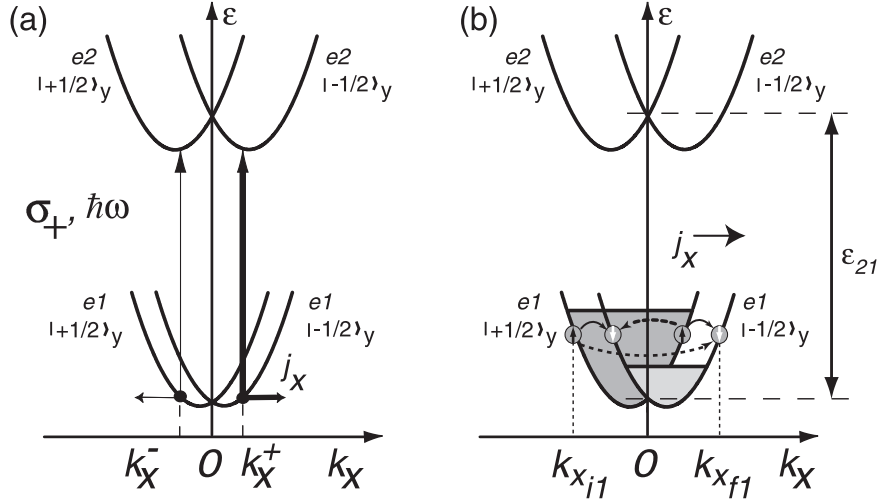}}

\caption{Microscopic picture of (a) circular photogalvanic effect
and (b)  spin-galvanic effect at inter-subband excitation in
C$_{2v}$ point group samples. In (a) the current $j_x$ is caused
by the imbalance of optical transition probabilities at $k^-_x$
and $k^+_x$ decaying with the momentum relaxation time $\tau_p$.
Excitation with $\sigma_+$ radiation of $\hbar \omega$ less than
the energy subband separation at $\bm{k}$=0, $\varepsilon_{21}$,
induces direct spin-conserving transitions (vertical arrows) at
$k_x^-$ and $k_x^+$. The rates of these transitions  are different
as illustrated by the different thicknesses of the arrows
(reversing the angle of incidence mirrors the  transition rates).
This leads to a photocurrent due to an asymmetrical distribution
of carriers in {\boldmath $k$}-space if the splitting of the $e1$
and $e2$ subbands is non-equal. Increase of the photon energy
shifts more intensive transitions to the left and less intensive
to the right resulting in a current sign change. In (b) the
current occurs after  thermalization in the lowest subband which
results in the spin orientation in the $e1$ subband. This
spin-galvanic current is caused by asymmetric spin-flip
scattering. The rate of spin-flip scattering depends on the value
of the initial and final $\bm{k}$-vectors.  Thus transitions
sketched by dashed arrows yield an asymmetric occupation of both
subbands and hence a current flow which decays with the spin
relaxation time $\tau_s$. The magnitude of the spin polarization
and hence the current depends on the initial absorption strength
but not on the momentum $\bm{k}$ of the transition. Therefore the
shape of the spectrum of the spin-galvanic current  follows the
absorption. } \label{fig3}
\end{figure}

It has been shown in~\cite{Nature02,PRL01} that in
quantum wells belonging to one of the gyrotropic crystal classes a
non-equilibrium spin polarization of electrons uniformly
distributed in space causes a directed motion of electrons in the
plane of the QW. On a microscopic level spin photocurrents are the
result of spin orientation in systems with {\boldmath $k$}-linear
terms in the electron effective Hamiltonian which are characteristic of gyrotropic
media. In general, two mechanisms contribute to spin
photocurrents: photoexcitation and scattering of photoexcited
carriers.  The first effect is the spin orientation induced
circular photogalvanic effect (CPGE) which is caused by an
asymmetry of the momentum distribution of  carriers excited in
optical transitions~\cite{PRL01,PRB03inv}. The second  effect is
the spin-galvanic effect which  in general does not need
optical excitation but is a result  of an asymmetric spin
relaxation~\cite{Nature02}. The current due to CPGE  is
spin polarized and decays with the momentum relaxation time
$\tau_p$ of photoexcited free carriers whereas the spin-galvanic
effect induced current is unpolarized but decays with the spin
relaxation time $\tau_s$. Both effects are illustrated in
Fig.~\ref{fig3}.

The change of sign of the photocurrent with  photon energy is
characteristic for CPGE at resonant transitions in $n$-type QWs
and has been described previously~\cite{PRB03inv}. As illustrated
in Fig.~\ref{fig3}a for $\sigma_+$ radiation  and at a small
photon energy less than the energy separation between $e1$ and
$e2$ at $k_x =0$, excitations occur preferentially at positive
$k_x$. We note that for $C_{2v}$ symmetry the optical transitions
are spin-conserving but spin-dependent~\cite{PRB03inv}. This
causes a stronger reduction in the electron population at positive
$k_x$ in the lower $\left|-1/2\right\rangle_y$-subband and
therefore a spin-polarized current in positive  $x$ direction. We
note that there is a corresponding  increase of the electron
population in the $e2$ $\left|-1/2\right\rangle_y$-subband, also
asymmetrical, but in our case this randomizes quickly via optical
phonon scattering and therefore does not contribute significantly
to the current~\cite{PRB03inv}. Increase of the photon energy
shifts the dominating transition towards negative $k_x$ and
reverses the current. In fact it has been shown that the CPGE at
inter-subband absorption in $n$-type QWs is proportional to the
derivative of the absorption spectrum~\cite{PRB03inv}. This
behaviour is observed for the current in $x$ $\parallel$
[1$\bar{1}$0] direction. In particular, the position of the sign
inversion of the current coincides with the maximum of the
absorption spectrum which shows that the spin-galvanic effect for
this direction is vanishingly small and the current is caused by
the CPGE.

In contrast to the CPGE the sign of the spin-galvanic current
is independent of the wavelength~\cite{PASPS02monop}. This can be
seen from Fig.~\ref{fig3}b which illustrates the origin of the
spin-galvanic effect. All that is required is a spin orientation
of the lower subband, and asymmetrical spin relaxation then drives
a current~\cite{Nature02}. In our case the spin orientation is
generated by resonant spin-selective optical excitation followed
by spin-non-specific thermalization. The magnitude of the spin polarization and
hence the current depends on the initial absorption strength but
not on the momentum $\bm{k}$ of transition. Therefore there is no
sign change and the shape of the spectrum follows the
absorption~\cite{PASPS02monop}. The lack of a sign change for
current along $y$ $\parallel$ [110]  in the experiment shows that
the spin-galvanic dominates for this orientation.

In order to understand the difference  between the two
orientations we now introduce a phenomenological picture for the
$C_{2v}$ symmetry representing samples investigated here.
Phenomenologically the spin-galvanic effect (SGE) and the circular
photogalvanic effect in $x$ and $y$ directions are given by
\begin{equation} j_{SGE,x} = Q_{xy} S_{y} \:,\: \: \: \: \: \: \: \: \: \: \:
j_{SGE,y} =  Q_{yx} S_{x}.
\label{eq19}
\end{equation}
\begin{equation}
j_{CPGE,x} = \gamma_{xy} \hat{e}_y E^2_0  P_{circ}  \:,\: \: \: \: \: \: \: \: \: \: \:
j_{CPGE,y} = \gamma_{yx} \hat{e}_x E^2_0  P_{circ}.
\label{eq11}
\end{equation}
where {\boldmath$j$} is the  photocurrent   density,
{\boldmath$Q$} and {\boldmath$\gamma$} are second rank
pseudo-tensors, {\boldmath$S$} is the average spin of electrons in
QWs, $E_0$, $P_{circ}$ and {\boldmath$\hat{e}$} are the amplitude
of the electromagnetic wave, the degree of circular polarization
and the unit vector pointing in the direction of light
propagation, respectively. In the present case {\boldmath$S$} is
obtained by  optical orientation, its sign and magnitude are
proportional to $P_{circ}$ and it is oriented along the in-plane
component of {\boldmath$\hat{e}$} (see Fig.~\ref{fig1}). Because
of  tensor equivalence of {\boldmath$Q$} and {\boldmath$\gamma$}
the spin-galvanic current induced by circularly polarized light
always occurs simultaneously with the CPGE. If the in-plane
component of {\boldmath$\hat{e}$}  is oriented along
 [1$\bar{1}$0] or [110], i.e. $x$ or $y$,
then both  currents flow normal to the light propagation direction.
The strength of the current is different for the radiation propagating
along $x$ or $y$. This is due to the non-equivalence of the
crystallographic axes [1$\bar{1}$0] and [110] because of the
two-fold rotation axis in $C_{2v}$ symmetry.

\begin{figure}
   \centerline{\epsfxsize 90mm \epsfbox{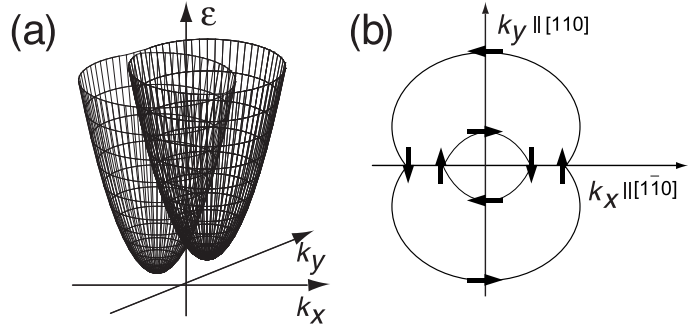}}

\caption{Schematic 2D  band structure with $\bm{k}$-linear terms
for $C_{2v}$ symmetry. The energy $\varepsilon$ is plotted as a
function of $k_x$ and $k_y$ for equal strength of the BIA and SIA
terms in the Hamiltonian. The bottom plate shows the distribution
of spin orientations at the 2D Fermi energy: (b) for BIA and SIA
terms with equal strength and (c) for different strength of SIA
and BIA terms  in the Hamiltonian. The difference of subband
spin-splitting in $x$ and $y$ directions, which is clearly seen
from (b) and (c) sketches, occurs due to non-equality of
$\beta_{xy} \sigma_x k_y$ and $\beta_{yx} \sigma_y k_x$. These
component of {\boldmath$\beta$}  may also be written as
$\beta_{xy} = (\beta_{BIA} +\beta_{SIA})/2$ and $\beta_{yx} =
(\beta_{BIA} - \beta_{SIA})/2$, respectively. Arrows indicate the
orientation of spins. } \label{fig4}
\end{figure}

Both currents are caused by  spin splitting of subbands in the
$\bm{k}$-space~\cite{PRL01,Nature02}. This splitting is due to
$\bm{k}$-linear terms in the Hamiltonian of the form
$\hat{H}^\prime=\sum_{lm}\beta_{lm}\sigma_l k_m$, where
$\beta_{lm}$ is a second rank pseudo-tensor and $\sigma_l$ are the
Pauli-matrices. The tensors {\boldmath$\gamma$}  and
{\boldmath$Q$} determining the current are related to the
transposed pseudo-tensor {\boldmath$\beta$}. They are subjected to
the same symmetry restrictions so that their irreducible
components  differ only by  scalar factors. The non-zero
components of the pseudo-tensor $\beta_{lm}$ depend on the
symmetry and the coordinate system used. For
(001)-crystallographic orientation grown QWs of $C_{2v}$ symmetry
and  in the coordinate system $(xyz)$, relevant to our
experimental set-up,  there are two non-zero  tensor elements
$\beta_{xy} \neq \beta_{yx}$ which may also be different for $e1$-
and $e2$ subbands. It is reasonable to introduce symmetric and
anti-symmetric tensor components $\beta_{BIA}^{(\nu)} =
(\beta_{xy}^{(\nu)} +\beta_{yx}^{(\nu)})/2$ and
$\beta_{SIA}^{(\nu)} = (\beta_{xy}^{(\nu)}
-\beta_{yx}^{(\nu)})/2$, where  $\nu$=1,2 indicates the $e1$ and
$e2$ subbands respectively. Here  $\beta_{BIA}^{(\nu)} $ and
$\beta_{SIA}^{(\nu)} $ result from bulk inversion asymmetry (BIA)
also called the Dresselhaus term~\cite{Dyakonov86p110} (including
a possible interface inversion asymmetry~\cite{Krebs96p1829}) and
from structural inversion asymmetry (SIA) usually called the
Rashba term~\cite{Bychkov84p78}, respectively.  In order to
illustrate band structures with a $\bm{k}$-linear term in
Fig.~\ref{fig4} we plotted  the energy $\varepsilon$ as a function
of $k_x$ and $k_y$ and constant energy surfaces for different
relations between $\beta_{BIA}$ and $\beta_{SIA}$ which are
assumed to be positive. The non-equivalence of $x$ and $y$
directions for $|\beta_{xy}| \neq |\beta_{yx}|$ is clearly seen
from Fig.~\ref{fig4}c.

As discussed above and sketched in Fig.~\ref{fig3} both CPGE and
spin-galvanic currents, say in $x$  direction, are caused by the
band splitting in $k_x$ direction and therefore are proportional
to $\beta_{yx}$ (for current in $y$-direction one should
interchange the indices $x$ and $y$ ). Then the currents in the
$x$ and $y$ directions read
\begin{equation} \label{biasia}
j_x =  A_{CPGE} [ (\beta^{(1)}_{BIA} - \beta^{(1)}_{SIA}) -
 (\beta^{(2)}_{BIA} - \beta^{(2)}_{SIA} ) ] P_{circ} \hat{e}_y
+ A_{SGE}( \beta^{(1)}_{BIA} -  \beta^{(1)}_{SIA} ) S_y
\end{equation}
and
\begin{equation} \label{biasia2}
j_y = A_{CPGE} [ (\beta^{(1)}_{BIA} + \beta^{(1)}_{SIA}) -
 (\beta^{(2)}_{BIA} + \beta^{(2)}_{SIA} ) ] P_{circ} \hat{e}_x
+ A_{SGE}( \beta^{(1)}_{BIA} + \beta^{(1)}_{SIA} ) S_x \:\:,
\end{equation}
where $A_{CPGE}$ and $A_{SGE}$ are factors related to
{\boldmath$\gamma$} and {\boldmath$Q$}, respectively. The
magnitude of the CPGE is determined by the value of $\bm{k}$ in
the initial and final states, and hence on the spin splitting
($\beta_{BIA}$ and $\beta_{SIA}$) of both $e1$ and $e2$ subbands.
In contrast, the spin-galvanic effect is due to relaxation between
the spin states of the lowest subband $e1$ and hence only on
$\beta^{(1)}_{BIA}$ and $\beta^{(1)}_{SIA}$. The
Eqs.~(\ref{biasia}) and (\ref{biasia2})  show that in directions
$x$ and $y$ the spin-galvanic effect and the CPGE are proportional
to terms containing the difference  and the sum, respectively, of
BIA and SIA terms. When they add (see Eq.~\ref{biasia2}) it
appears in our samples that the spin-galvanic effect  dominates
over the CPGE which is proved by the lack of sign change for
currents along the $y$ direction in Fig.~\ref{fig2}. Conversely
when BIA and SIA terms subtract (see Eq.~\ref{biasia}) the
spin-galvanic effect is suppressed and the CPGE dominates. We
would like to emphasize at this point that at the frequency where
CPGE is equal to zero for both directions, the current obtained is
caused by the spin-galvanic effect  only.

The occurrence of a spin-galvanic current  is due to the spin
dependence of the electron scattering matrix elements $M_{
\bm{k}^\prime \bm{k}}$. The 2 $\times$ 2 matrix $\hat{M}_{
\bm{k}^\prime \bm{k}}$ can be written as a linear combination of
the unit matrix $\hat{I}$ and Pauli matrices as follows
\begin{equation}
\hat{M}_{ \bm{k}^\prime \bm{k}  } = A_{ \bm{k}^\prime \bm{k}  }
\hat{I} + \mbox{\boldmath$ \sigma$} \cdot \mbox{\boldmath$ B$}_{
\bm{k}^\prime \bm{k} } \:,
\label{matelis}
\end{equation}
where $A^*_{ \bm{k}^\prime \bm{k} } =A_{ \bm{k} \bm{k}^\prime}$,
$B^*_{ \bm{k}^\prime \bm{k}  } = B_{ \bm{k} \bm{k}^\prime}$ due to
hermiticity of the interaction and $A_{- \bm{k}^\prime, - \bm{k} }
=A_{ \bm{k} \bm{k}^\prime}$, $B_{- \bm{k}^\prime, - \bm{k}  } = -
B_{ \bm{k} \bm{k}^\prime}$ due to the symmetry under time
inversion. The spin-dependent part of the scattering amplitude
in (001)-grown QW structures is given by~\cite{Averkiev02pR271}
\begin{equation}
\mbox{\boldmath$ \sigma$} \cdot \mbox{\boldmath$ B$}_{
\bm{k}^\prime \bm{k}  } = v( \bm{k}  - \bm{k}^\prime) [ \sigma_x
(k^\prime_y + k_y) - \sigma_y (k^\prime_x + k_x)] \:.
\label{matelis1}
\end{equation}
where  $v$($\bm{k}$ -- $\bm{k}$$^\prime)$ is a function defined
in~\cite{Averkiev02pR271}. We note that Eq.~(\ref{matelis1})
determines the spin relaxation time, $\tau^\prime_s$, due to the
Elliot-Yafet mechanism. Then, for instance, for the spin component
$S_x$ assuming a Boltzmann distribution, the spin-galvanic current
in $y$ direction has the form
\begin{equation} \label{sgey}
j_{SGE,y} =  \frac{4 \pi e}{m^*} S_x
\sum_{\tilde{\bm{k}}\,\tilde{\bm{k}}'} \left( \tilde{k}'_y -
\tilde{k}_y \right) \left( \tilde{k}'_x + \tilde{k}_x \right)^2
\left|\,v(\tilde{\bm{k}}-\tilde{\bm{k}}'-2\bm{k}_0)\right|^2
\tau_p
\end{equation}
$$ \times \, f\left( \frac{\hbar^2 \tilde{\bm{k}}^2}{2 m^*}
\right) \delta \left( \frac{\hbar^2 \tilde{\bm{k}}'^2}{2 m^*} -
\frac{\hbar^2 \tilde{\bm{k}}^2}{2 m^*} \right) $$
where $e$ is the electron charge, $\tau_p$ is the momentum
scattering time, $f$ is the distribution function, $\delta$ is the
delta function, $m^*$ is the electron effective mass,
$\tilde{{\bm{k}}} = {\bm{k}} + {\bm{k}}_0$,
$\tilde{{\bm{k^\prime}}} = {\bm{k^\prime}} - {\bm{k}}_0$, and
$\bm{k}_0=(m^* \beta_{xy} /\hbar^2,0,0)$. By using
Eq.~(\ref{sgey}) we can estimate the spin-galvanic current as
\begin{equation} \label{est1}
j_{SGE,y}= Q_{yx}S_x \propto e\: n_e \frac{\beta_{xy}^{(1)}}{\hbar}
\frac{\tau_p }{\tau^\prime_s} S_x
\:.
\end{equation}
Since scattering is the origin of the spin-galvanic effect,  the
spin-galvanic current, $j_{SGE}$, is determined by the
Elliot-Yafet spin relaxation time. The relaxation time
$\tau^\prime_s$ is proportional to the momentum relaxation time
$\tau_p$.  Therefore the ratio $\tau_p / \tau_s^\prime$ in
Eq.~(\ref{est1}) does not depend on the momentum relaxation time.
The in-plane average spin $S_x$ in Eq.~(\ref{est1}) decays  with
the total spin relaxation time $\tau_s$ (which may have a
contribution from any spin relaxing process). Thus the time decay
of the spin-galvanic current following the pulsed photoexcitation
is determined by $\tau_s$. The current in $x$ direction may be
obtained by exchanging $x$ and $y$ in Eq.~(\ref{est1}).

For the present case, where spin relaxation is obtained as a
result of  inter-subband absorption of circularly polarized
radiation, the current is given by
\begin{equation}
j_{SGE,x} \sim e\:\frac{\beta_{yx}^{(1)}}{\hbar}
\frac{\tau_p \tau_s}{\tau^\prime_s} \frac{\eta_{21} I}{\hbar \omega}
 P_{circ} \xi \hat{e}_y \:,\: \: \: \: \: \: \: \: \: \: \:
j_{SGE,y} \sim e\:\frac{\beta_{xy}^{(1)}}{\hbar}
\frac{\tau_p \tau_s}{\tau^\prime_s} \frac{\eta_{21} I}{\hbar \omega}
 P_{circ} \xi \hat{e}_x \:.
\label{j_sge}
\end{equation}
$\eta_{21}$ is the absorbance at the transitions  between $e1$
 and $e2$ subbands, $I$ is the radiation intensity. The
parameter $\xi$ varying between 0 and 1 is the ratio of
photoexcited electrons relaxing to the $e1$ subband with and
without spin-flip. It determines the degree of spin polarization
in the lowest subband (see Fig.~\ref{fig3}b) and depends on the
details of the relaxation mechanism. Optical orientation requires
$\xi\neq 0$~\cite{Meier,Parson71p1850,IT}. Eqs.~(\ref{j_sge}) show
that the  spin-galvanic current is proportional to the absorbance
and is determined by the spin splitting in the first subband,
$\beta^{(1)}_{yx}$ or $\beta^{(1)}_{xy}$.

In conclusion we observed the spin-galvanic effect under
all-optical excitation and without applying external magnetic
fields by making use of the interplay of the Rashba and
Dresselhaus splitting of the conduction band. Our results
demonstrate in a direct way the non-equivalence of the [110] and
[1$\bar{1}$0] directions in zinc-blende structure QWs. The results
also clearly show the difference between the microscopic pictures
for spin-galvanic and CPGE, effects which have the same
phenomenological description.

We thank L.\,E. Golub for many helpful discussions. Financial
support from the DFG, the RFBR,  INTAS, the EPSRC (UK) and FOM
(NL) is gratefully acknowledged. The authors are grateful to the
FELIX facility team and especially for the skillful assistance of
Dr.\,A.\,F.\,G.~van\,der\,Meer.

\end{document}